# Low-Dimensional High-Fidelity Kinetic Models for NO$_X$ Formation by a Compute Intensification Method

Mark Kelly[a*], Harry Dunne[a], Gilles Bourque[b,c], Stephen Dooley[a]

[a]*School of Physics, Trinity College Dublin, Dublin, Ireland*
[b]*Siemens Energy Canada Ltd, Montréal, QC H9P 1A5, Canada*
[c]*McGill University, Montréal, QC H3A 0G4, Canada*

___

**Abstract**

A novel compute intensification methodology to the construction of low-dimensional, high-fidelity "compact" kinetic models for NO$_X$ formation is designed and demonstrated. The method adapts the data intensive Machine Learned Optimization of Chemical Kinetics (MLOCK) algorithm for compact model generation by the use of a Latin Square method for virtual reaction network generation. A set of logical rules are defined which construct a minimally sized virtual reaction network comprising three additional nodes (N, NO, NO$_2$). This NO$_X$ virtual reaction network is appended to a pre-existing compact model for methane combustion comprising fifteen nodes.

The resulting eighteen node virtual reaction network is processed by the MLOCK coded algorithm to produce a plethora of compact model candidates for NO$_X$ formation during methane combustion. MLOCK automatically; populates the terms of the virtual reaction network with candidate inputs; measures the success of the resulting compact model candidates (in reproducing a broad set of gas turbine industry-defined performance targets); selects regions of input parameters space showing models of best performance; refines the input parameters to give better performance; and makes an ultimate selection of the best performing model or models.

By this method, it is shown that a number of compact model candidates exist that show fidelities in excess of 75% in reproducing industry defined performance targets, with one model valid to >75% across fuel/air equivalence ratios of 0.5-1.0. However, to meet the full fuel/air equivalence ratio performance envelope defined by industry, we show that with this minimal virtual reaction network, two further compact models are required.

*Keywords:* Compact Kinetic Models, Mechanism Optimisation, Machine Learning, Gas-Turbine Physics

___

*Corresponding author.

# 1. Introduction

Nitrogen oxides ($NO_X$) are harmful emissions legally regulated across the world due to their negative effects on human health and the wider environment. As a result, power generating combustion devices, such as gas turbine technologies, must advance through increasing power-emission efficiencies. This demand is concurrent with ongoing operational needs for gas turbines to provide clean and efficient, low $NO_X$ combustion for a wide range of fuel choices. To find the burner geometry and operating conditions that maximises the emission efficiency, a greater understanding of the combustion phenomena occurring inside the combustor is needed. Computational fluid dynamics (CFD) reacting flow simulations can be performed rapidly at operating conditions where physical experimentation is both expensive, slow, and challenging to perform. A key component in reacting flow simulations is an accurate chemical kinetic model. Detailed chemical kinetic models describe the actual combustion reaction mechanism at fundamental detail and generally reproduce combustion kinetic phenomena to quantitative accuracy. However, the computational cost of a simulation scales logarithmically with the number of chemical species contained in a kinetic model. Thus, using this level of detail in CFD is computationally prohibitive.

To allow for deployment of multi-dimensional reacting flow simulations in rapid, iterative computational design of combustors, a kinetic model of much reduced dimensionality is needed. This challenge has been the purpose of mechanism reduction research, which is now a mature field, where the state-of-art and its limits is widely recognised [6-8]. In this regard, our state of art is given by models produced through graphing techniques such as *Directed Relation Graph (DRG)* [10] or *Path Flux Analysis (PFA)* [9]. These are physics-based algorithms that quantify the importance of each species in the reaction network component of the model to the high fidelity replication of sets of selected information resulting from calculations of the detailed model. Usually, the information targets are species concentrations, pressure vs. time, or heat release behaviours.

However, mechanism reduction methodologies are intrinsically limited, because as species are removed from the reaction network, the fidelity of the calculations decreases until a critical point of necessary detail is reached. After this point, further removal of species leads to a deterioration in the model performance. Figure 1 shows this limit to be 25 species for the example of the methane system. The limit is due to the "over-reduction" of the reaction network, removing species that are important in mediating the reaction flux through the system from reactants to products. The removal of this important information from the reaction network results in a severe compromising of the authentic reaction flux. This threshold is thus a limit condition due to the nature of modelling approaches that seek to describe the underlying phenomenology at a level of detail that is authentic to fundamental physics. One can thus expect this limit to be general across all conventional mechanism reduction techniques. However, for application of chemical reaction kinetics in multidimensional CFD, industry demand in the design of low-$NO_X$ combustors is for kinetic models to comprise fewer than approximately twenty species.

# 2. Compact Kinetic Models

The concept of a Compact Kinetic Model has the potential to meet this challenge. In a compact model, all detail that is not absolutely necessary to the calculation of a defined set of performances is removed. The errors that result owing to the removal of this detail are compensated for by optimisation of any of the reaction kinetic, molecular thermodynamic or mass/energy transport terms comprising the model. In this way, the model

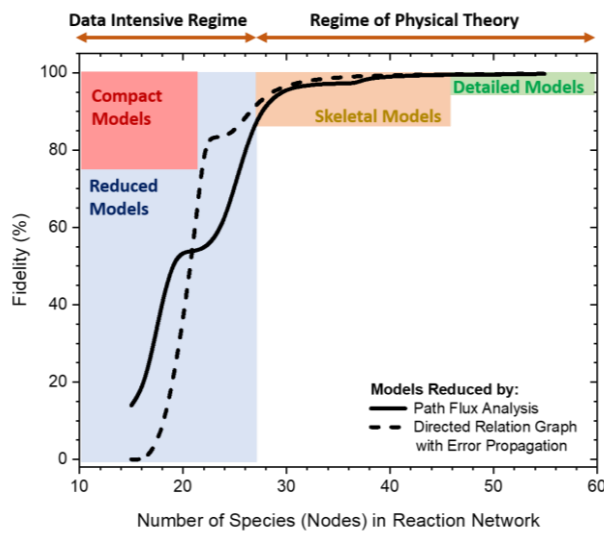

Figure 1. Dependence of model calculation fidelity for methane combustion on complexity of reaction network, as indicated by number of species (or nodes), where Detailed, Skeletal, Reduced and Compact Models occupy different regions of detail-fidelity space. A series of reduced models are produced by the Directed Relation Graph with Error Propagation reduction method [1, 2] from NUIGMECH1.0 [3] via DoctorSMOKE++ [4, 5], and by the Path Flux Analysis method [9]. Fidelity assessment was performed using ignition delay time (i.e., constant UV) calculations at: 1-40 atm, 1100-2000 K, methane/air mixture fractions 0.5 – 1.5.

description of the reacting flow is *virtual* rather than attempting to be physically authentic.

The FAIR [11] principles of data science require that computational models be interoperable, meaning they can be readily used with current software infrastructure ("plug and play"). As most commercial combustion modelling platforms allow the ChemKin format of chemical reaction and transport model description, it is thus a logical starting point to test if high fidelity compact models can be produced in this format. This imposition therefore excludes the use of mathematical descriptors that are not directly compatible with the ChemKin format. It also excludes other implementations that require either modification of the solver or additional sub-routines to be exercised (e.g., quasi-steady state approximation). This is obviously a severe restriction. Despite this, studies applying optimization techniques of this type to the task of finding the ideal set of kinetic descriptors for severely reduced kinetic models have been appearing. They include, genetic algorithms approaches [12], gradient based algorithm approaches [13], techniques emphasising artificial neural networks [14, 15] and particle swarm algorithms [16], each showing varying degrees of success. And so, success or failure in the creation of high-fidelity minimally-sized FAIR kinetic models by data intensive approaches presents an important go/no-go turning point for an advanced description of combustion reaction kinetics.

The compact model concepts interfaces the fundamental physics-based description of reaction kinetics with a less physical, data-science perspective through the four archetypal kinetic model components:

1. *A Virtual Reaction Network* in place of the reaction mechanism. The purpose of the virtual reaction network is to provide minimal, but sufficient, degrees of freedom to the reacting flux such that certain combinations of virtual reaction rate constants can exist that yield models of high fidelity to the set of reaction kinetic target properties defined by the user. Formally, the virtual reaction network is comprised of nodes (species) joined together by connections (virtual reactions) of varying weights (virtual reaction rate constants). The configuration of each is adjustable to a possibly inexhaustible number of combinations with the objective of replicating some defined set of combustion kinetic performance attributes. The more complex or higher accuracy sought in performance attributes (e.g. species fractions, combustion phenomena), and the wider the range of performance conditions (e.g. mixture fraction, temperature, pressure, combustion phenomena), the more complex the virtual reaction network needed.

2. *Virtual Reaction Rate Constants (or Weights)* define the magnitude of the connection between the nodes in the virtual reaction network. Analogous to the elementary reaction rate constants, but not describing of a physically real chemical reaction, units are $cm^3mol^{-1}s^{-1}$.
3. *Molecular Thermodynamic Descriptors* including the standard state molar enthalpy (H) and entropy (S), and the constant pressure heat capacity ($C_P$) of each species are retained from the detailed kinetic model.
4. *Mass and Energy Transport Descriptors* of each species are retained from the detailed kinetic model.

## 3. Methodology: Compact Models for $NO_X$ Formation in Methane Combustion

MLOCK1.0 [17] is a coded semi-autonomous compaction algorithm that manipulates components 1 and 2 to perform the compaction. MLOCK1.0 creates virtual reaction networks by exercising conventional model reduction procedures and then optimises the reaction kinetics through systematic perturbation of virtual reaction rate constants of important connections. For this study, MLOCK1.0 is adapted to create MLOCK1.1 as described below, toward the challenge of minimalistic $NO_X$ modelling.

*3.1 Virtual Reaction Network Generation*

In the case of $NO_X$ chemistry, the actually physically occurring reaction mechanism is relatively complex, and consequently, the detailed kinetic models that simulate this already comprise over fifteen species, e.g., Grimech3.0 [18] has seventeen nitrogen-containing species. Notwithstanding this, the methodology employed here recognises the industry-defined performance envelope and necessarily aims to describe this complex chemical kinetics with minimalistic detail.

Industry demands a reaction network comprising as few species additional to the fuel species as possible. For $NO_X$, by definition this is two (NO and $NO_2$). For this minimalistic challenge, from experience we know that the implementation of conventional mechanism reduction as deployed in MLOCK1.0 would not yield minimally sized reaction networks. Instead, for this study, a virtual reaction network of minimum complexity was constructed from combinatoric theory using a Latin Square design, as shown in Figure 2. This approach allows for the systematic generation of the maximum number of connections for a given set of nodes (species array). Thus, the minimum sized virtual reaction network with sufficient degrees of freedom for successful optimisation should hypothetically be created. If a successful optimisation is not achieved, the species array may be iteratively expanded providing increased degrees of freedom to the optimization algorithm, whilst still retaining the minimal sized reaction network that is viable.

The Latin Square reaction network generation method allows for the nitrogen containing nodes to interact with the nodes of the associated fuel reaction network, by systematic combinatorics [19]. Given this, and that the MLOCK optimisation algorithm allows vast degrees of freedom, it is important that safe-guarding measures are imposed to prevent adulteration of the reaction flux of the base-fuel combustion kinetics. Therefore, two selection criteria are applied to define the node array for the Latin Square:

1. Nodes must contain elements that comprise the target nodes i.e., O or N for NO or $NO_2$,
2. Nodes must not be a molecule (excluding combustion products and reactants i.e., $CO_2$, $H_2O$ NO, $NO_2$ and $O_2$).

This results in a nine node Latin Square given by Figure 2. To generate the connections the following rules are then applied to the Latin Square:

3. Reactant combinations must only be two-node.
4. All nodes in the underlying fuel sub-model are permitted to be a product node of a connection.
5. Connection products may be one-, two- and three-node.

The virtual reaction network construction is then completed by applying the below rules to the results of the Latin Square:

6. All connections are set as reversible.
7. All duplicate connections are removed.
8. Connections that are found to cause any considerable numerical stiffness in optimisation are set as unidirectional and subsequently removed from the network if stiffness persists.

This procedure generates a virtual reaction network, comprising three nodes (N, NO, $NO_2$) interacting through 29 connections, or which two are unidirectional. The remaining challenge is to assign to the connections a set of weights (virtual reaction rate constants), that allow the overall compact model to achieve the performance properties defined by industry to high fidelity.

*3.2 Optimisation of Kinetic Component (MLOCK1.1)*

Performance optimisation is performed by MLOCK1.1 which systematically; generates; assigns; and evaluates sets of

Table 1. Gas-turbine industry defined performance targets for $NO_X$ formation during methane combustion. For PSR and flame, only calculations at 10 atm and at selected equivalence ratios are used to construct the OEFs (i.e., training data), as described in text. The remaining conditions are unseen to the compact model candidates and are thus considered as validation. An overall model fidelity of >75% is deemed acceptable.

| Reactor Physics | $NO_X$ Property | Base Fuel Property | Pressure | Fuel/Air Mixture Fraction | Temperature | Use |
|---|---|---|---|---|---|---|
| Perfectly Stirred | [NO], [$NO_2$] | [CO], [OH] | 1 – 40 atm | 0.4 – 1.4 | 1100 – 2000 K | Training & Validation |
| Ignition Delay Time (Constant UV) | [NO], [$NO_2$] | IDT, [CO], [OH] | 1 – 40 atm | 0.4 – 1.4 | 1100 – 2000 K | Validation |
| Freely Propagating Flame | [NO], [$NO_2$] | [CO], Laminar Flame Speed | 1 – 40 atm | 0.4 – 1.4 | 473 – 673 K | Training & Validation |

| | $CO_2$ | $H_2O$ | $HO_2$ | $O$ | $O_2$ | $OH$ | $N_2$ | $NO$ | $NO_2$ | $N$ |
|---|---|---|---|---|---|---|---|---|---|---|
| $CO_2$ | | | | | | | $NO + CO + N$ | $NO_2 + CO$ | | $NO + CO$ |
| $H_2O$ | | | | | | | $NO + N + H_2$ | $NO_2 + H_2$ $NO_2 + H + H$ | $NO + H_2O_2$ $NO + H_2 + O_2$ $NO + OH + OH$ $NO + HO_2 + H$ | $NO + H_2$ $NO + H + H$ |
| $HO_2$ | | | | | | | $NO_2 + N + H$ $NO + N + OH$ $NO + NO + H$ | $NO_2 + OH$ $NO_2 + O + H$ $N + O_2 + OH$ | $NO + H_2O_2 + O$ $NO + H_2O + O_2$ | $NO + OH$ $NO + O + H$ $NO_2 + H$ |
| $O$ | | | | | | | $NO + N$ $N + N + O$ $NO + NO$ $NO_2 + N$ $NO + N + O$ | $NO_2$ $N + O_2$ | $NO + O_2$ | $NO$ |
| $O_2$ | | | | | | | | | | $NO + O$ $NO_2$ |
| $OH$ | | | | | | | $NO + N + H$ | $NO_2 + H$ $N + HO_2$ | $NO + HO_2$ $NO + H + O_2$ | $NO + H$ |
| $N_2$ | $NO + CO + N$ | $NO + N + H_2$ | $NO_2 + N + H$ $NO + N + OH$ $NO + NO + H$ | $NO + N$ $N + N + O$ $NO_2 + OH$ | $NO + NO$ $NO_2 + N$ $NO + N + O$ | $NO + N + H$ | | | $NO + NO + N$ | |
| $NO$ | $NO_2 + CO$ | $NO_2 + H_2$ $NO2 + H + H$ | $NO_2 + H_2$ $NO_2 + O + H$ $N + O_2 + OH$ | $NO_2$ | $NO_2 + O$ | $NO_2 + H$ | | | | $N_2 + O$ |
| $NO_2$ | | $NO + H_2O_2$ $NO + H_2 + O_2$ $NO + OH + OH$ $NO + HO_2 + H$ | $NO + H_2O_2 + O$ $NO + H_2O + O_2$ | $NO + O_2$ | | $NO + HO_2$ $NO + OH + O$ $NO + H + O_2$ | $NO + NO + N$ | $N_2 + O_2$ $N + NO_2$ | | $NO + NO$ $N_2 + O_2$ |
| $N$ | $NO + CO$ | $NO + H_2$ $NO + H + H$ | $NO + OH$ $NO + O + H$ $NO_2 + H$ | $NO$ | $NO + O$ $NO_2$ | $NO + H$ | | $N_2 + O$ | $NO + NO$ $N_2 + O_2$ | $N_2$ |

Legend:
- Hydrocarbon Array
- First Combinatorial Array
- Second Combinatorial Array
- Removed due to stiffness.
- Removed by rate of production analysis.
- Retained in final network.

Figure 2. Latin Square used to construct the $NO_X$ virtual reaction network. The second combinatorial array produces 93 connections. 52 duplicates are removed (Step 7). By processing, MLOCK sets 2 connections as unidirectional and removes a further 12 due to numerical stiffness (Step 8). Rate of production analysis removes a further 18 connections from the final compact model construct. This results in 11 discreet connections in the virtual reaction network of the final compact model construct.

Arrhenius parameters for each connection in the $NO_X$ virtual reaction network. Figure 3 shows the MLOCK1.1 process flow to consist of two "*Scans*" that minimise objective error functions (OEFs) through the generation and evaluation of 80,000 compact model candidates by systematic simultaneous perturbation of all 29 virtual reaction rate constants ($k_v$) (before ROP removal). This is performed by random combination of each of the $A$, $n$, and $E_A$ terms of the standard ChemKin format, with subsequent logarithmically uniform sampling of the virtual reaction rate constant ($k_v$) library obtained. These scans search an area of the entire multi-dimensional virtual reaction rate constant bound space so as to assess the suitability of various regions in producing compact models that replicate industry performance targets to high fidelity. Table 1 details the industry defined performance targets used in this work, defining a useful model to show an overall fidelity of at least 75%.

*Objective Error Functions:* The MLOCK method relies on the generation and calculation of Objective Error Functions (OEFs) to quantify and objectify the performance of each compact model candidate. In this study, the OEFs are comprised of data produced by exercising the NUIGMECH1.0 detailed model for natural gas combustion, including $NO_X$ formation. In this way, the compact models produced by this study are a "compaction" of the NUIGMECH1.0 detailed model for $NO_X$ formation during methane combustion.

*0-D OEF & Course Scan:* First, the Coarse Scan employs PSR simulations to emphasise a kinetically driven reaction system, where, as in conventional model reduction, an accurate replication of PSR calculations is a good indication that the pertinent kinetics are accurately described. For this assessment, a zero-dimensional objective error function (0-D OEF) is constructed. The 0-D OEF is composed of NUIGMECH1.0 calculations of NO and $NO_2$ mole fractions in a perfectly stirred reactor (PSR) at 1100, 1500 and 2000 K, at 1, 10 and 40 atm; and at equivalence ratios 0.7, 0.9, 1, 1.1 and 1.4. The 0-D OEF is constructed such that the entire time-evolved fidelity of the model calculation is evaluated, as given by Equation 1:

$$0\text{-}D\ OEF = \frac{1}{N_J N_K} \sum_k \sum_j \sum_i \left( w_j \frac{\left| [X_i^{red}(t_j)]_k - [X_i^{det}(t_j)]_k \right|}{X_{i,k}^{det}(t_j)} \right) \quad (1)$$

where, $[X_i^{red}(t_j)]_k$ refers to the mole fraction of the $i$'th species at time point $j$, at the $k$'th set of temperature, pressure, and equivalence ratio conditions, as calculated by the compact model candidate. $t_j$ corresponds to the time at which NUIGMECH1.0 calculation of $[X_i]$ is equal to 100%, 70%, 50% and 30% the maximum mole fraction at that condition, i.e., $t(n_n\ [X_i^{det}]_{max})$ where n is 1, 0.7, 0.5 and 0.3. $w_j$ is the $j^{\prime th}$ weighting factor (1, 0.7, 0.5, and 0.3) assigned to each of the time points to assign more importance to errors at times closer to the maximum mole fractions of NO and NO$_2$.

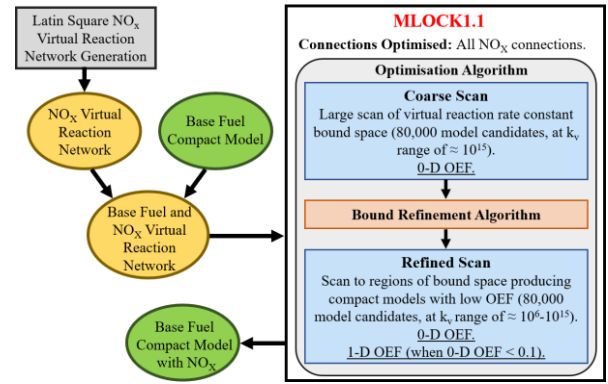

Figure 3. Process flow for MLOCK1.1. For the models discussed in this study, note that (i) the 0-D (OEF) used in both the Coarse and Refined Scans is composed of NUIGMECH1.0 perfectly stirred reactor calculations of NO and $NO_2$ mole fractions at conditions of 1100 – 2000 K, 1 – 40 atm and fuel/air equivalence ratios of 0.7 – 1.4. (ii) the 1-D OEF used in the Refined Scan is composed of NUIGMECH1.0 premixed laminar flame calculations of NO and $NO_2$ mole fractions at conditions of 573 K, 10 atm and fuel/air equivalence ratios of 0.6 – 1.4.

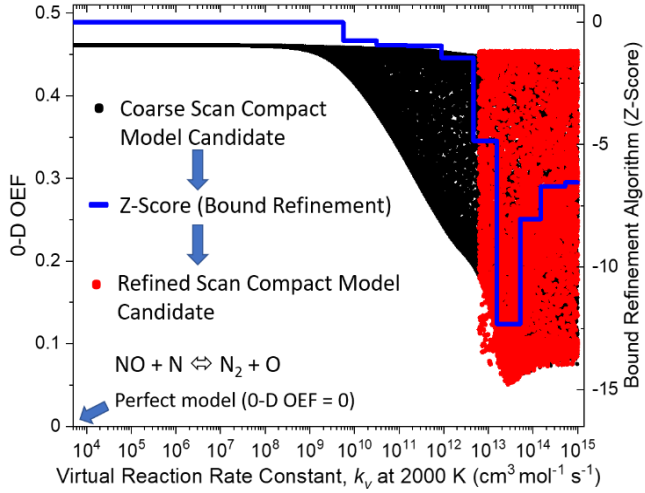

Figure 4. 0-D OEF (Objective Error Function) of compact model candidates produced in the Coarse and Refined Scans as a function of the virtual reaction rate constant of the connection NO + N <=> N$_2$ + O. The z-score is calculated by the bound refinement algorithm following the Coarse Scan and directs the search algorithm to regions of the virtual reaction rate constant bound space that produce models with low 0-D OEF. 2000 K is a nominal reference.

From Figure 3, MLOCK 1.1 initiates with a large coarse grid-like scan to generate an approximate representation of the parameter bound space, searching a range in $k_v$ of approximately fifteen orders of magnitude by creating 80,000 compact model candidates at logarithmically uniform intervals of $k_v$. For each compact model candidate, the 0-D OEF is calculated and compiled into a report.

*Bound Refinement Algorithm, Z-Score:* The information in this report is automatically analysed by a bound refinement algorithm to score each region of the $k_v$ bound space on the likelihood of a good model being produced in that region. The bound refinement algorithm operates on the principal of a binomial test in which the null hypothesis is that the percentage of compact model candidates in a region of $k_v$ that have an OEF less than a certain value, is the same as the percentage across the whole population (full $k_v$ range). A Z-test is then performed to test if the null hypothesis is true or false for each region of $k_v$.

First, the value of the OEF that corresponds to the 99'th percentile of the 80,000 compact model candidates produced in the Coarse Scan is calculated and the number of compact model candidates in the complete population with an OEF below this value is recorded. The $k_v$ bound space is then divided into a series of regions, and the number of models in each region that have an OEF below this threshold value is counted. If the proportion of models below the OEF threshold in the full population is $\bar{p}$, and within a region of $k_v$ it is found to be p, then for a region of $k_v$ with n compact model candidates, the Z-score for that region is,

$$Z = \frac{\bar{p} - p}{\sqrt{\frac{\bar{p}(1-\bar{p})}{n}}} \quad (3)$$

Lower Z-values correspond to a greater proportion of low OEF compact model candidates produced in the specific $k_v$ region. The algorithm then outputs regions of $k_v$ with a Z-score below a certain value.

*Refined Scan, 1-D OEF:* This information is subsequently used by the Refined Scan to constrain the search algorithm to areas of $k_v$ bound space identified by the bound refinement algorithm as producing a large proportion of high-fidelity models. This increases the resolution of these promising regions of $k_v$ bound space, increasing the algorithm's efficiency, and leading to the finding of more, and higher fidelity compact model candidates. If no values of $k_v$ are outputted from the refinement algorithm, this indicates a weak coupling between any value of $k_v$ and the OEF for a given connection. In this scenario, the bounds are kept at those used in the Coarse Scan. This process is performed independently for every connection undergoing optimisation.

In the Refined Scan, compact model candidates with low 0-D OEF are tested against 1-D calculations through a 1-D OEF. The 1-D OEF quantifies the error between NO and NO$_2$ mole fractions in a freely propagating premixed flame as calculated by each compact model candidate and the NUIGMECH1.0 detailed model. The 1-D OEF is evaluated using six positions within the flame, such that the error in the unburned gas, reaction zone and post-flame regions are assessed. This is performed at 10 atm, 573 K, and equivalence ratios 0.6, 0.8, 1.0, 1.2, and 1.4 by Equation 2:

$$1\text{-}D\ OEF = \frac{1}{N_n N_K} \sum_k \sum_x \sum_i \left( \frac{\left| [X_i^{red}(x_n)]_k - [X_i^{det}(x_n)]_k \right|}{X_{i,k}^{det}(x_n)} \right) \quad (2)$$

where $[X_i^{red}(x_n)]_k$ refers to the mole fraction of the $i^{\prime th}$ species at position *n*, at the $k^{\prime th}$ set of temperature, pressure, and equivalence ratio conditions, as calculated by the compact model candidate.

This Coarse Scan – Bound Refinement – Refined Scan process is demonstrated in Figure 4 for the example of the $k_v$ – fidelity behaviour of the connection N + NO ⇔ N$_2$ + O. A strong coupling between the value of $k_v$ and performance of the compact model candidates is evident.

*Compact Model Selection:* Following the refined scan, the compact model candidate with the lowest average OEF (i.e., (0-D OEF + 1-D OEF)/2) is selected as the final Compact Model.

Finally, MLOCK1.1 performs a rate of production analysis to identify and remove connections making negligible contribution to the NO$_X$ virtual reaction flux, which can thus be removed without affecting the compact model's calculations. This leads to the removal of 18 connections (black in Figure 2). This procedure results in a virtual reaction network comprised of 18 nodes and 71 connections. The unseen performance of the best compact model candidates of this architecture are accounted below, and all of the best performing models are available at [20].

## 4. Results and Discussion

To wholly quantify model accuracies, Fidelity Indexes of the form of Equation 4 are defined to simply sum the relative errors of each point in the overall industry-defined performance set,

$$Fidelity\ (\%) = \frac{100}{N} \sum_k \left( 1 - \frac{|X_k^{compact} - X_k^{detailed}|}{X_k^{detailed}} \right) \quad (4)$$

where $X_k$ is the species mole fraction, ignition delay time (IDT), or laminar flame speed at the k'th set of pressure, temperature and equivalence ratio conditions and N is the number of conditions included.

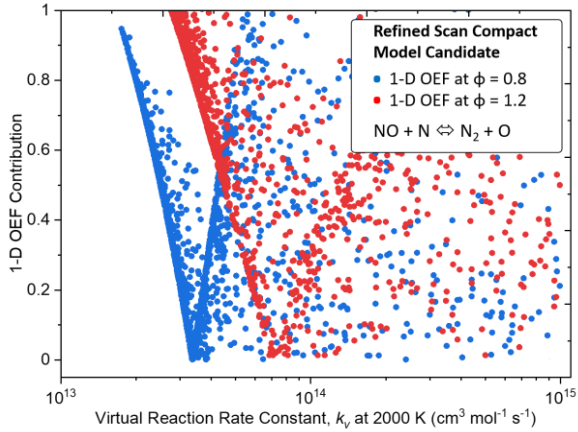

Figure 5. 1-D OEF (Objective Error Function) at lean (blue) and rich (red) fuel/air mixture fractions of compact model candidates produced in the Refined Scan as a function of the virtual reaction rate constant of the connection NO + N $\Leftrightarrow$ N$_2$ + O.

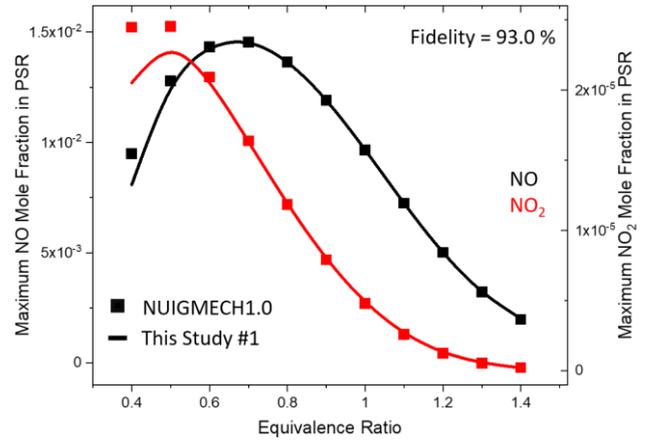

Figure 6. Maximum NO and NO$_2$ mole fractions from perfectly stirred reactor calculations using NUIGMECH1.0 (symbol) and the compact model produced in this work (line) at 10 atm and 1500 K at lean to rich fuel/air mixture fractions.

First, an essential result is that the addition of the NO$_X$ sub-model to the base fuel sub-model does not perturb the calculations of the underlying base fuel sub-model. The supplemental material provides comprehensive evidence demonstrating this.

The NO$_X$ performance target parameter space is also vast, and so only representative calculations can be shown here, with full accounting in the supplemental material. Table 2 summarises the overall performance in this regard. Table 2 shows that compact models with this virtual reaction network of the desired >75% fidelity for the entire equivalence ratio range, can be found for the 0-D set of performance conditions, but not for the 1-D set of performance conditions. Table 2 also summarises the performance of the best performing models at select ranges of equivalence ratio. Here, we show that three different models (comprising the same virtual reaction network, but different sets of weights) are needed to attain >75% fidelity across the entire equivalence ratio range.

The Z-score significance rating of the coarse scan 0-D OEFs (see SM) shows that this is so, due to the limited complexity available to a NO$_X$ virtual reaction network that is three-node. The Z-score of the virtual reaction NO + N $\Leftrightarrow$ N$_2$ + O (that of Figure 4) is -12.32. That of the next largest virtual reaction is -3.01, indicating a probability that the other virtual reactions exhibit a much weaker influence on the system. As MLOCK perturbs the magnitude of $k_v$ of all the connections simultaneously, this means that combinations of $k_v$ of all the other 28 connections (prior to ROP removal) cannot be found to show a magnitude of influence similar to singular perturbation of this term.

This behaviour is further explained by analysing Figure 5, which shows the 1-D OEF at lean (0.8) and rich (1.2) equivalence ratios of compact model candidates produced in the refined scan as a function of $k_v$ for the connection NO + N $\Leftrightarrow$ NO + O. To produce a model that performs well at lean conditions (blue symbols), the value of $k_v$ for this connection must be approximately 3.5 x10$^{13}$ cm$^3$ mol$^{-1}$ s$^{-1}$. However, this value for $k_v$ results in a poor replication of NO$_X$ (large OEFs) at rich conditions (red symbol). Consequently, a compact model that has fidelity >57% at a comprehensive range of equivalence ratios cannot be produced using the minimal three-node NO$_X$ virtual reaction network.

Therefore, with this minimal construct one can either; (i) produce a compact model that has excellent fidelity at a limited range of equivalence ratios, or (ii) produce a compact model with a relatively lower, but more consistent fidelity across the entire equivalence ratio range. Compact models satisfying both options are presented by this study, with their performance summarised by Table 2. A third option is to expand the node array of the Latin Square to increase the degrees of freedom available to the optimisation algorithm. It is likely that this option will result in models of higher fidelity over broader ranges of the performance space but will also result in a larger computational cost. This study presents compact models satisfying option (i). Note that from Figure 5, there are some model candidates with values of $k_v$ greater than $10^{14}$ where it may appear that a compact model has been produced that performs well at both lean and rich conditions for the same value of $k_v$. However, upon further inspection, it is found that these models either do not actually perform well (i.e., an inaccuracy in the OEF) or that the model performs well at these specific equivalence ratios but poorly elsewhere.

Table 2. Compact model performance summary and comparison to State-of-Art. The fidelity of all model calculations are evaluated at 1 – 20 atm. Flame calculations were performed at equivalence ratios 0.4 – 1.5 Ignition Delay Time (IDT) calculations are performed at 1100 – 2000 K & equivalence ratios 0.5 – 1.5. PSR calculations are performed at 1100 – 2000 K & equivalence ratios 0.4 – 1.5. Note: The Sun et al. model does not contain NO$_2$, hence, NO$_X$ fidelity for this model is simply NO fidelity. NO$_X$ Fidelity calculated at range of equivalence ratios specified in table.

| Model | No. of Nodes (Species) | Equivalence Ratios Evaluated | IDT Fidelity (%) | NO$_X$ Fidelity in PSR (%) | Laminar Flame Speed Fidelity (%) | NO$_X$ Fidelity in Flame (%) |
|---|---|---|---|---|---|---|
| *Compact Models for Performance Option (i)* | | | | | | |
| This Study #1 for $\phi$ 0.4-1.0 | 18 | 0.5, 0.6, 0.7, 0.8, 0.9, 1.0 | 87.1 | 91.2 | 88.6 | 84.3 |
| This Study #2 for $\phi$ 0.4 & 1.4 | 18 | 0.4 & 1.4 | 87.1 | 86.4 | 88.6 | 76.5 |
| This Study #3 for $\phi$ 1.1-1.3 | 18 | 1.1, 1.2 & 1.3 | 87.1 | 85.0 | 88.6 | 78.6 |
| *Compact Models for Performance Option (ii)* | | | | | | |
| Sun et al. [24] | 22 | 0.4 – 1.4 | 58.0 | 77.1 | 83.2 | 61.9 |
| This Study #1 for $\phi$ 0.4-1.4 | 18 | 0.4 – 1.4 | 87.1 | 89.5 | 88.6 | 56.9 |

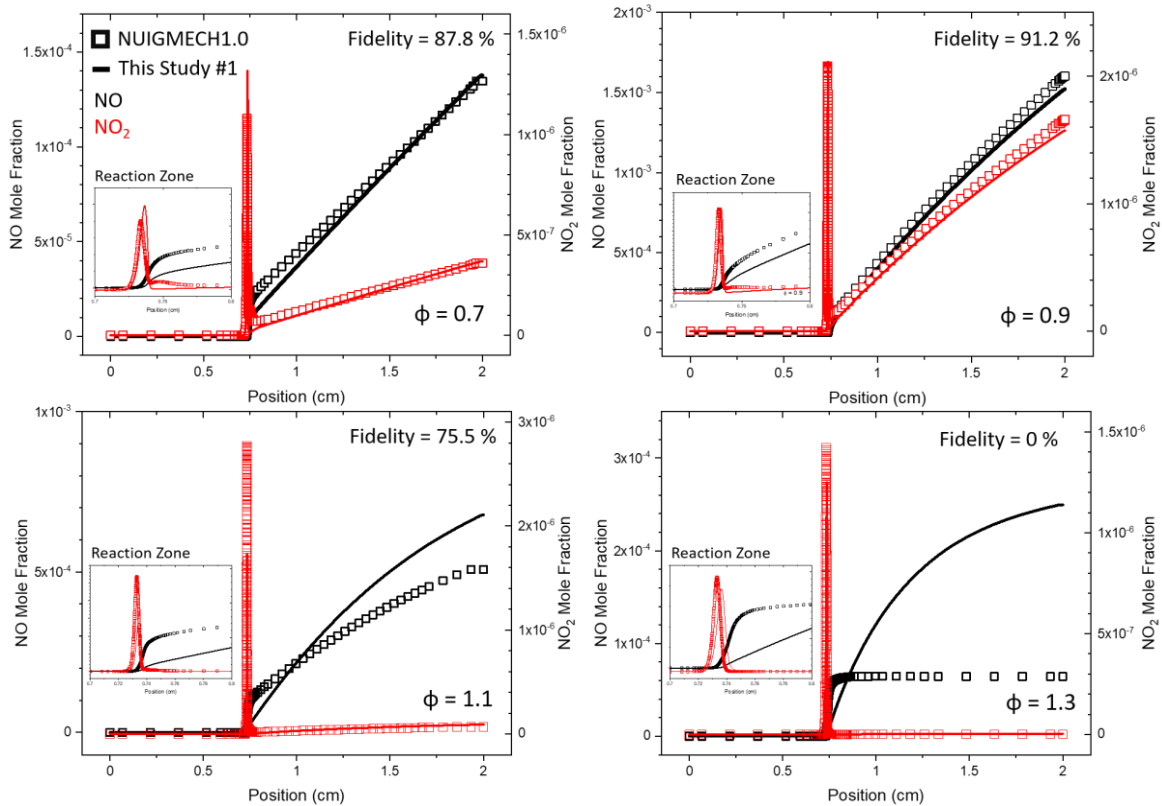

Figure 7. NO and $NO_2$ mole fractions in laminar premixed flame calculations using NUIGMECH1.0 (symbol) and compact model #1 produced in this study (line) at 573 K, 10 atm and fuel/air equivalence ratios 0.7 – 1.3. INSET: Expanded view of reaction zone.

The seen and unseen performance of each of the compact models #1, #2 and #3 for $NO_X$ formation during methane combustion as proposed by this study is provided as Supplemental Material. Figures 6 and 7 provide representative performance of Compact Model #1, which performs best over the widest range of equivalence ratios (option (i)).

*4.2 State-of-Art Comparison*

The challenge of describing $NO_X$ formation using a minimally-sized kinetic model has been attempted before. However, most works apply analytical techniques that are not compatible with industry standard reaction kinetic solvers such as ChemKin and/or are not presented in the standard ChemKin format e.g., [21-23]. Thus, such descriptions may not abide by the FAIR principles of data science discussed earlier. However, recently Sun et al. [24] report a 22 species model for methane/air combustion with NO (but not $NO_2$) formation. To our knowledge, this is the only methane + $NO_X$ compact model of the order of approximately twenty species that uses standard ChemKin reaction kinetics descriptors. Sun et al. appear to apply the Direct Relation Graph with Sensitivity Analysis method to reduce GRIMech3.0 conventionally. An important difference to the MLOCK1.1 methodology of this work, is that the Sun et al. method considers interactions of the base fuel reaction kinetics with the $NO_X$ reaction kinetics. That is, the prompt $NO_X$ pathways are retained in the compact model. This is at the cost of four additional species, relative to the MLOCK1.1 minimalistic combinatoric method of this work. Note that the Sun et al. model also comprises just three $NO_X$ species (N, NNH, NO). The fidelity of the compact models of this work, and that of Sun et al. is compared in Figure 8 with respect to $NO_X$ mole fractions in a premixed laminar flame at 1-20 atm, which is the pressure range the Sun et al. model was developed for. Note, the model produced in the work of Sun et al. does not include $NO_2$, but this is a small fraction of the total $NO_X$ formed at all conditions. From Figure 8, the Sun et al. model shares the inability of the compact models of this work to show high fidelity at all conditions, also suggesting an apparent limitation to a three-node $NO_X$ virtual reaction network. A more detailed summary of the performance of both models is provided in Table 2. The compact model of this work, Model #1, shows high fidelity under purely kinetic PSR conditions, but loses fidelity at flame conditions, due to poor performance at the extreme lean and rich equivalence ratios. The Sun et al. model shows similar behaviour but is of significantly lower fidelity at PSR conditions.

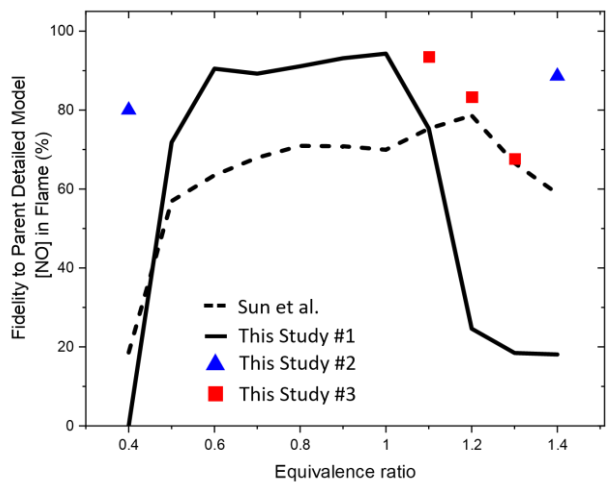

Figure 8. Fidelity of models produced in this work and in the work of Sun et al. to respective parent detailed model calculations of NO mole fractions in laminar premixed flames at 1 – 20 atm. Note that both studies show relative consistency in an apparent limit of $NO_X$ virtual reaction networks that's are three-node. Other combinations of weights can yield compact models of improved fidelity (symbols).

## 5. Conclusions

A novel data-orientated compute intensification methodology to the construction of low-dimensional high-fidelity "compact" kinetic models for $NO_X$ formation is designed and demonstrated. A detail-fidelity cost-benefit performance envelope for modelling of $NO_X$ formation in methane combustion is defined for gas-turbine industry interests. The MLOCK coded algorithm for compact model generation is adapted to $NO_X$ modelling with the addition of a Latin Square combinatoric method to allow creation of a virtual reaction network of minimal complexity with just three nitrogen containing nodes (N, NO & $NO_2$). With this construct, MLOCK1.1 automatically generates and tests 160,000 compact model candidates. Of these, for kinetically driven PSR behaviours it is shown that several models show fidelities > 80% across the entire range of performance conditions. However, for flame conditions, it is not possible to find one model of > 57% fidelity across the entire industry-defined performance envelope. It is shown that the loss of fidelity is due to the distinctive $NO_X$ kinetic behaviours at extreme lean, and extreme rich equivalence ratios, which cannot be simultaneously captured by such a simple three-node virtual reaction network. However, using the same minimal three-node virtual reaction network, it is shown that three distinct compact models, each valid at a limited range of equivalence ratios do collectively meet the industry defined performance envelope. Though further development is required, the results of this work do appear to support the concept that high-fidelity, low-dimensional "compact" models can be automatically produced using data-intensive methodologies, with the output conforming to FAIR industry standard (i.e. ChemKin) model formats.

## Acknowledgements

The research reported in this publication was support by funding from Siemens Canada Limited and the Sustainable Energy and Fuel Efficiency (SEFE) Spoke of MaREI, the SFI Centre for Energy, Climate and Marine Research (16/SP/3829). MK is supported by the Irish Research Council (GOIPG/2020/1043). Calculations were performed on the Boyle cluster maintained by the Trinity Centre for High Performance Computing.